\batchmode
\documentstyle[12pt]{article}
\newcommand{\nc}{\newcommand}
\nc{\renc}{\renewcommand}

\nc{\half}{{\textstyle{1\over2}}}
\nc{\etal}{\mbox{\it et al. }}
\nc{\ie}{{\it i.e.}}
\nc{\eg}{{\it e.g.}}

\renc{\thefootnote}{\arabic{footnote}}
\nc{\capt}[1]{{\bf Figure.} {\small\sl #1}}

\nc{\eqs}[2]{\mbox{Eqs.~(\ref{#1},\,\ref{#2})}}
\nc{\eq}[1]{\mbox{Eq.~(\ref{#1})}}

\nc{\figs}[2]{\mbox{Figs.~(\ref{#1},\,\ref{#2})}}
\nc{\fig}[1]{\mbox{Fig~.(\ref{#1})}}

\nc{\tag}[1]{\label{#1} \marginpar{{\footnotesize #1}}}
\nc{\mtag}[1]{\label{#1} \mbox{\marginpar{{\footnotesize #1}}}}
\renc{\baselinestretch}{1.5}
\newlength{\overeqskip}
\newlength{\undereqskip}
\nc{\be}[1]{\begin{equation} \mbox{$\label{#1}$}}
\nc{\bea}[1]{\begin{eqnarray} \mbox{$\label{#1}$}}
\nc{\Section}[2]{\section{#2}\label{#1}}
\nc{\Bibitem}[1]{\bibitem{#1}}
\nc{\Label}[1]{\label{#1}}

\nc{\eea}{\vspace{\undereqskip}\end{eqnarray}}
\nc{\ee}{\vspace{\undereqskip}\end{equation}}
\nc{\bdm}{\begin{displaymath}}
\nc{\edm}{\end{displaymath}}
\nc{\dpsty}{\displaystyle}
\nc{\bc}{\begin{center}}
\nc{\ec}{\end{center}}
\nc{\ba}{\begin{array}}
\nc{\ea}{\end{array}}
\nc{\bab}{\begin{abstract}}
\nc{\eab}{\end{abstract}}
\nc{\btab}{\begin{tabular}}
\nc{\etab}{\end{tabular}}
\nc{\bit}{\begin{itemize}}
\nc{\eit}{\end{itemize}}
\nc{\ben}{\begin{enumerate}}
\nc{\een}{\end{enumerate}}
\nc{\bfig}{\begin{figure}}
\nc{\efig}{\end{figure}}
\nc{\arreq}{&\!=\!&}
\nc{\arrmi}{&\!-\!&}
\nc{\arrpl}{&\!+\!&}
\nc{\arrap}{&\!\!\!\approx\!\!\!&}
\nc{\non}{\nonumber\\*}
\nc{\align}{\!\!\!\!\!\!\!\!&&}

\def\lsim{\; \raise0.3ex\hbox{$<$\kern-0.75em
      \raise-1.1ex\hbox{$\sim$}}\; }
\def\gsim{\; \raise0.3ex\hbox{$>$\kern-0.75em
      \raise-1.1ex\hbox{$\sim$}}\; }
\nc{\DOT}{\hspace{-0.08in}{\bf .}\hspace{0.1in}}
\nc{\Laada}{\hbox {$\sqcap$ \kern -1em $\sqcup$}}
\nc\loota{{\scriptstyle\sqcap\kern-0.55em\hbox{$\scriptstyle\sqcup$}}}
\nc\Loota{{\sqcap\kern-0.65em\hbox{$\sqcup$}}}
\nc\laada{\Loota}
\nc{\qed}{\hskip 3em \hbox{\BOX} \vskip 2ex}

\nc{\real}{{\rm I \! R}}
\nc{\Z}{{\sf Z \!\!\! Z}}
\nc{\complex}{{\rm C\!\!\! {\sf I}\,\,}}
\def\bigid{\leavevmode\hbox{\small1\kern-3.8pt\normalsize1}}
\def\id{\leavevmode\hbox{\small1\kern-3.3pt\normalsize1}}
\nc{\slask}{\!\!\!/}
\nc{\bis}{{\prime\prime}}
\nc{\pa}{\partial}
\nc{\na}{\nabla}
\nc{\ra}{\rangle}
\nc{\la}{\langle}
\nc{\goto}{\rightarrow}
\nc{\swap}{\leftrightarrow}

\nc{\EE}[1]{ \mbox{$\cdot10^{#1}$} }
\nc{\abs}[1]{\left|#1\right|}
\nc{\at}[2]{\left.#1\right|_{#2}}
\nc{\norm}[1]{\|#1\|}
\nc{\abscut}[2]{\Abs{#1}_{\scriptscriptstyle#2}}
\nc{\vek}[1]{{\rm\bf #1}}
\nc{\integral}[2]{\int\limits_{#1}^{#2}}
\nc{\inv}[1]{\frac{1}{#1}}
\nc{\dd}[2]{{{\partial #1}\over{\partial #2}}}
\nc{\ddd}[2]{{{{\partial}^2 #1}\over{\partial {#2}^2}}}
\nc{\dddd}[3]{{{{\partial}^2 #1}\over
	{\partial #2 \partial #3}}}
\nc{\dder}[2]{{{d #1}\over{d #2}}}
\nc{\ddder}[2]{{{d^2 #1}\over{d {#2}^2}}}
\nc{\dddder}[3]{{d^2 #1}\over
	{d #2 d #3}}
\nc{\dx}[1]{d\,^{#1}x}
\nc{\dy}[1]{d\,^{#1}y}
\nc{\dz}[1]{d\,^{#1}z}
\nc{\dl}[1]{\frac{d\,^{#1}l}{(2\pi)^{#1}}}
\nc{\dk}[1]{\frac{d\,^{#1}k}{(2\pi)^{#1}}}
\nc{\dq}[1]{\frac{d\,^{#1}q}{(2\pi)^{#1}}}

\nc{\cc}{\mbox{$c.c.$ }}
\nc{\hc}{\mbox{$h.c.$ }}
\nc{\cf}{cf.\ }
\nc{\erfc}{{\rm erfc}}
\nc{\Tr}{{\rm Tr\,}}
\nc{\tr}{{\rm tr\,}}
\nc{\pol}{{\rm pol}}
\nc{\sign}{{\rm sign}}
\nc{\bfT}{{\bf T }}

\def\GeV{{\rm\ GeV}}
\def\MeV{{\rm\ MeV}}

\def\TeV{{\rm\ TeV}}

\nc{\cA}{{\cal A}}
\nc{\cB}{{\cal B}}
\nc{\cD}{{\cal D}}
\nc{\cE}{{\cal E}}
\nc{\cG}{{\cal G}}
\nc{\cH}{{\cal H}}
\nc{\cL}{{\cal L}}
\nc{\cO}{{\cal O}}
\nc{\cT}{{\cal T}}
\nc{\cN}{{\cal N}}
\nc{\rvac}[1]{|{\cal O}#1\rangle}
\nc{\lvac}[1]{\langle{\cal O}#1|}
\nc{\rvacb}[1]{|{\cal O}_\beta #1\rangle}
\nc{\lvacb}[1]{\langle{\cal O}_\beta #1 |}
\nc{\bb}{\bar{\beta}}
\nc{\bt}{\tilde{\beta}}
\nc{\ctH}{\tilde{\cal H}}
\nc{\chH}{\hat{\cal H}}
\nc{\1}{\aa}
\nc{\2}{\"{a}}
\nc{\3}{\"{o}}
\nc{\4}{\AA}
\nc{\5}{\"{A}}
\nc{\6}{\"{O}}
\nc{\al}{\alpha}
\nc{\g}{\gamma}
\nc{\Del}{\Delta}
\nc{\e}{\epsilon}
\nc{\eps}{\epsilon}
\nc{\lam}{\lambda}
\nc{\om}{\omega}
\nc{\Om}{\Omega}
\nc{\ve}{\varepsilon}
\nc{\mn}{{\mu\nu}}
\nc{\k}{\kappa}
\nc{\vp}{\varphi}

\nc{\advp}[3]{{\it  Adv.\ in\ Phys.\ }{{\bf #1} {(#2)} {#3}}}
\nc{\annp}[3]{{\it  Ann.\ Phys.\ (N.Y.)\ }{{\bf #1} {(#2)} {#3}}}
\nc{\apl}[3]{{\it  Appl. Phys. Lett. }{{\bf #1} {(#2)} {#3}}}
\nc{\apj}[3]{{\it  Ap.\ J.\ }{{\bf #1} {(#2)} {#3}}}
\nc{\apjl}[3]{{\it  Ap.\ J.\ Lett.\ }{{\bf #1} {(#2)} {#3}}}
\nc{\app}[3]{{\it Astropart.\ Phys.\ }{{\bf #1} {(#2)} {#3}}}
\nc{\cmp}[3]{{\it  Comm.\ Math.\ Phys.\ }{{ \bf #1} {(#2)} {#3}}}
\nc{\cqg}[3]{{\it  Class.\ Quant.\ Grav.\ }{{\bf #1} {(#2)} {#3}}}
\nc{\epl}[3]{{\it  Europhys.\ Lett.\ }{{\bf #1} {(#2)} {#3}}}
\nc{\ijmp}[3]{{\it Int.\ J.\ Mod.\ Phys.\ }{{\bf #1} {(#2)} {#3}}}
\nc{\ijtp}[3]{{\it Int.\ J.\ Theor.\ Phys.\ }{{\bf #1} {(#2)} {#3}}}
\nc{\jmp}[3]{{\it  J.\ Math.\ Phys.\ }{{ \bf #1} {(#2)} {#3}}}
\nc{\jpa}[3]{{\it  J.\ Phys.\ A\ }{{\bf #1} {(#2)} {#3}}}
\nc{\jpc}[3]{{\it  J.\ Phys.\ C\ }{{\bf #1} {(#2)} {#3}}}
\nc{\jap}[3]{{\it J.\ Appl.\ Phys.\ }{{\bf #1} {(#2)} {#3}}}
\nc{\jpsj}[3]{{\it J.\ Phys.\ Soc.\ Japan\ }{{\bf #1} {(#2)} {#3}}}
\nc{\lmp}[3]{{\it Lett.\ Math.\ Phys.\ }{{\bf #1} {(#2)} {#3}}}
\nc{\mpl}[3]{{\it  Mod.\ Phys.\ Lett.\ }{{\bf #1} {(#2)} {#3}}}
\nc{\ncim}[3]{{\it  Nuov.\ Cim.\ }{{\bf #1} {(#2)} {#3}}}
\nc{\np}[3]{{\it  Nucl.\ Phys.\ }{{\bf #1} {(#2)} {#3}}}
\nc{\npps}[3]{{\it  Nucl.\ Phys.\ Proc.\ Suppl.\ }{{\bf #1} {(#2)} {#3}}}
\nc{\pr}[3]{{\it Phys.\ Rev.\ }{{\bf #1} {(#2)} {#3}}}
\nc{\pra}[3]{{\it  Phys.\ Rev.\ A\ }{{\bf #1} {(#2)} {#3}}}
\nc{\prb}[3]{{\it  Phys.\ Rev.\ B\ }{{{\bf #1} {(#2)} {#3}}}}
\nc{\prc}[3]{{\it  Phys.\ Rev.\ C\ }{{\bf #1} {(#2)} {#3}}}
\nc{\prd}[3]{{\it  Phys.\ Rev.\ D\ }{{\bf #1} {(#2)} {#3}}}
\nc{\prl}[3]{{\it Phys.\ Rev.\ Lett.\ }{{\bf #1} {(#2)} {#3}}}
\nc{\pl}[3]{{\it  Phys.\ Lett.\ }{{\bf #1} {(#2)} {#3}}}
\nc{\prep}[3]{{\it Phys.\ Rep.\ }{{\bf #1} {(#2)} {#3}}}
\nc{\prsl}[3]{{\it Proc.\ R.\ Soc.\ London\ }{{\bf #1} {(#2)} {#3}}}
\nc{\ptp}[3]{{\it  Prog.\ Theor.\ Phys.\ }{{\bf #1} {(#2)} {#3}}}
\nc{\ptps}[3]{{\it  Prog\ Theor.\ Phys.\ suppl.\ }{{\bf #1} {(#2)} {#3}}}
\nc{\physa}[3]{{\it  Physica\ A\ }{{\bf #1} {(#2)} {#3}}}
\nc{\physb}[3]{{\it  Physica\ B\ }{{\bf #1} {(#2)} {#3}}}
\nc{\phys}[3]{{\it Physica\ }{{\bf #1} {(#2)} {#3}}}
\nc{\rmp}[3]{{\it  Rev.\ Mod.\ Phys.\ }{{\bf #1} {(#2)} {#3}}}
\nc{\rpp}[3]{{\it Rep.\ Prog.\ Phys.\ }{{\bf #1} {(#2)} {#3}}}
\nc{\sjnp}[3]{{\it Sov.\ J.\ Nucl.\ Phys.\ }{{\bf #1} {(#2)} {#3}}}
\nc{\spjetp}[3]{{\it Sov.\ Phys.\ JETP\ }{{\bf #1} {(#2)} {#3}}}
\nc{\yf}[3]{{\it Yad.\ Fiz.\ }{{\bf #1} {(#2)} {#3}}}
\nc{\zetp}[3]{{\it Zh.\ Eksp.\ Teor.\ Fiz.\  }{{\bf #1}  {(#2)} {#3}}}
\nc{\zp}[3]{{\it Z.\ Phys.\ }{{\bf #1} {(#2)} {#3}}}
\nc{\ibid}[3]{{\sl ibid.\ }{{\bf #1} {#2} {#3}}}
\nc{\rf}[1]{(\ref{#1})}
\nc{\nn}{\nonumber \\*}
\nc{\bfB}{\bf{B}}
\nc{\bfv}{\bf{v}}
\nc{\bfx}{\bf{x}}
\nc{\bfy}{\bf{y}}
\nc{\vx}{\vec{x}}
\nc{\vy}{\vec{y}}
\nc{\oB}{\overline{B}}
\nc{\oI}{\overline{I}}
\nc{\oR}{\overline{R}}
\nc{\rar}{\rightarrow}
\nc{\ti}{\times}
\nc{\slsh}{\hskip-5pt/}
\nc{\sm}{Standard~Model~}
\nc{\MP}{M_{\rm Pl}}
\nc{\tp}{t_{\rm Pl}}
\nc{\ave}{\bar{E}}

\nc{\eff}{{\rm eff}}
\nc{\kk}{\vek{k}}
\nc{\pp}{{\rm p}}
\nc{\ga}{g_{a\gamma}}
\nc{\vv}{\\}
\nc{\eee}{{\bf E}}
\nc{\bbb}{{\bf B}}
\nc{\qcd}{T_{\rm QCD}}
\nc{\G}{\rm \ G}
\def\vec#1{{\bf #1}}

\def\lae{\;^{<}_{\sim} \;} \def\gae{\; ^{>}_{\sim} \;} 

\def\ell{e^{c}LL}

\begin{document}
{\title{\vskip-2truecm{\hfill {{\small \\
	\hfill \\
	}}\vskip 1truecm}
{\LARGE Non-Thermal Dark Matter, High Energy Cosmic Rays
 and Late-Decaying Particles From Inflationary Quantum Fluctuations}}
{\author{
{\sc  John McDonald$^{1}$}\\
{\sl\small c$ \backslash $o Dept. of Physics and Astronomy, 
University of Glasgow, 
Glasgow G12 8QQ, SCOTLAND}
}
\maketitle
\begin{abstract}
\noindent

            It has been suggested that the origin
 of cosmic rays above the GZK limit might be
 explained by the decay of particles, $X$, with mass of the order of 
$10^{12} \GeV$. Generation of heavy particles from inflationary quantum
 fluctuations is a prime candidate for the origin of the 
decaying $X$ particles. 
 It has also been suggested that the problem of non-singular galactic 
halos might be
 explained if dark matter originates non-thermally from the decay of particles, 
$Y$, such that there is a free-streaming length of the order of 0.1Mpc. 
Here we explore the possibility that quantum fluctuations might 
account for the $Y$ particles as well as the $X$ particles. 
For the case of non-thermal WIMP dark matter
 with unsuppressed weak interactions we find that there
 is a general problem with deuterium photo-dissociation, 
disfavouring WIMP dark matter candidates. 
For the case of more general dark matter particles, which may
 have little or no interaction with conventional matter, we 
discuss the conditions under which $X$ and $Y$ scalars or fermions 
can account for non-thermal dark matter and cosmic rays. For the case where $X$ 
and $Y$ scalars are simultaneously produced, we show that galactic halos 
are likely to have a dynamically significant component of $X$ 
scalar cold dark matter 
 in addition to the dominant non-thermal dark matter component.

\end{abstract}
\vfil
\footnoterule
{\small $^1$mcdonald@physics.gla.ac.uk}

\thispagestyle{empty}
\newpage
\setcounter{page}{1}

\section{Introduction}

                  There have been a number of surprising observations
 made recently in cosmology. 
One has been the discrepency between simulations of
 galaxy halo formation
 and small-scale structure based on
 cold dark matter (CDM) and observations, which indicate that there
 is more small-scale
 structure than observed and that the halos are much less
 singular than predicted \cite{gal}. In particular, numerical simulations
 of CDM halo formation show that the abundance of dark matter subhalos
 and so dwarf galaxies within a galaxy should be the same as the abundance of 
subhalos within galactic clusters, whereas observations indicate that the
 abundance within galaxies is much less \cite{moore}.
 Various explainations have been
 put forward, including 
self-interacting dark matter \cite{sidm,ob}, halos made of condensates of
 ultra-light scalar  particles \cite{cond} and warm
 dark matter \cite{wdm}. Recently it has also been suggested
 that dark matter could
 originate non-thermally
 from heavy particle (or topological defect) decay \cite{ntdm}.  

           Another puzzle is the existence of
 ultra-high energy cosmic rays (UHECR) beyond the $10^{11} \GeV$
 GZK cut-off \cite{uhecr,sb}. A simple possible
 explaination of the UHECR observations
 would be to have a heavy decaying
 particle of mass $\approx 10^{12} \GeV$
 and lifetime longer than the age of the Universe, 
such that high energy proton, nuclei and photon 
 primaries can originate within the 50Mpc mean free path for 
energies greater than the GZK cut-off
 \cite{dp}. A natural origin for such heavy decaying particles is from
 quantum fluctuations generated during inflation \cite{grav}.

             In this paper we consider whether the decaying particles which may
 explain non-singular halos and those which account for UHECR 
could both originate from quantum fluctuations during inflation.  
The halo problem is then assumed to be
 solved by a density of non-thermal dark matter (NTDM) coming
 from the decay of heavy particles which we label
 $Y$, such that the non-singular halos are a result of a
 free-streaming length $\approx 0.1$Mpc
 \cite{ntdm}, whilst the UHECR are explained by
 decaying particles $X$ with masses of
 the order of $10^{12} \GeV$ \cite{sb}. 

          The paper is organized as follows. In section 2 we discuss
 the general conditions for solving the
 galactic halo problem via NTDM from heavy
 particle decay. We also briefly review the conditions required to 
account for UHECR via heavy particle decay.
 In section 3 we consider the special case of weakly interacting 
massive particle (WIMP) dark matter with 
unsuppressed weak interactions, in particular the conditions
required to evade energy loss of the
 WIMPs via scattering from the thermal background whilst preserving
primordial nucleosynthesis.  
In Section 4 we discuss whether the galactic halo problem and UHECR
 can be explained by $X$ and $Y$ particles generated 
during inflation and the
 possible observable consequences of this. In Section 5 we present
 our conclusions. In an Appendix we discuss the scattering
 cross-section and
 energy loss per scattering for a Majorana fermion scattering
 from thermal background particles. 

\section{General Conditions for NTDM and UHECR from
 Decaying Particles}

\subsection{Non-Thermal Dark Matter}

            The problem of singular galactic halos and excessive structure
 on small-scales 
can be solved if the dark matter has a
 free-streaming length of the order of 0.1Mpc \cite{ntdm}. 
To achieve this in the case of heavy dark matter particles, 
the dark matter must have a non-thermal
 origin, such as a decaying massive particle\footnote{An
 alternative is to have warm dark matter, a thermal distribution of 
particles of mass approximately 1 KeV \cite{wdm}.}.
Suppose the
 dark matter particles originate
 from decay of a massive particle at $t=t_{d}$. Then the
 co-moving free-streaming length is given by \cite{eu}
\be{e1} \lambda_{FS} =  \int_{a_{d}}^{a_{EQ}}
 \frac{v}{a^2H}  da 
\approx \frac{a_{NR}}{a_{EQ}^{2} H_{EQ}}
 \left(1 + ln \left(\frac{a_{EQ}}{a_{NR}}\right)
\right)     ~,\ee
where $a_{EQ}$ is the scale factor at matter-radiation
 equality, $a_{d}$ is the scale
 factor at $t_{d}$, $v$ is the dark matter particle
 velocity, and we have used the fact that
 $v \propto a^{-1}$ once the freely propagating
 dark matter particles are 
non-relativistic. 
If we assume that the initial dark matter particle energy is 
$\beta m_{Y}$, we obtain 
\be{e3} \lambda_{FS} \approx 
\frac{r_{c}}{a_{EQ}^{2} H_{EQ}} \left(1 + ln 
\left(\frac{a_{EQ}}{r_{c}}\right)
\right)   ~,\ee
where \cite{ntdm}
\be{e4}   r_{c} = \frac{\beta m_{Y} a_{d}}{m_{\chi}}      ~\ee
and $m_{\chi}$ is the dark matter particle mass. 
With $a=1$ at present, $r_{c}$ is equal to the present 
velocity of the dark matter particles. With $a_{EQ} = 
4.3 \times 10^{-5} 
(\Omega_{m} h^{2})^{-1} $, $\lambda_{FS} = 0.1$Mpc 
is obtained for \cite{ntdm}
\be{e5}   r_{c} \approx 2.4 \times 10^{-8} 
\left(\frac{\lambda_{FS}}{0.1 Mpc}\right)     ~.\ee
This is the condition for dark matter particles from $Y$ decay
 to account for non-singular
 galactic halos. 
Using $a_{d} = \left(\frac{g(T_{\gamma})}{g(T_{d})}\right)^{1/3}
\frac{T_{\gamma}}{T_{d}}$, 
this then fixes the $Y$ mass as a function of the decay temperature,
\be{e6}  m_{Y} \approx 4 \times 10^{7} \; T_{d}
\left(\frac{r_{c}}{10^{-7}\beta}\right)
\left(\frac{g(T_{d})}{g(T_{\gamma})}\right)^{1/3}
\left(\frac{m_{\chi}}{100 \GeV}\right)
 \GeV       ~,\ee
where $g(T)$ is the number of degrees of freedom in thermal
 equilibrium and 
$T_{\gamma}$ is the photon temperature at present. 

          In addition, the dark matter particle
 density, assumed to be dominated by $\chi$,
 must satisfy $\Omega_{\chi}
 \approx 0.3$. (We assume throughout a flat Universe with a
 cosmological constant $\Omega_{\Lambda} \approx 0.7$.)
Assuming that the number of dark matter particles from each
 $Y$ decay is $\epsilon$, we have  
\be{e7} \Omega_{\chi} = \frac{\epsilon m_{\chi} 
n_{Y}(T_{\gamma})}{\rho_{c}}       ~,\ee
where $\rho_{c}$ is the critical energy density and 
$n_{Y}(T_{\gamma}) = g(T_{\gamma}) T_{\gamma}^{3}/ g(T_{d})
T_{d}^{3}$ is the $Y$
 number density at present in the absence of $Y$ decays.
The requirement $\Omega_{\chi} \approx 0.3$
 will impose a constraint on the reheating 
temperature after inflation, $T_{R}$. 

\subsection{UHECR} 

              The observed UHECR are not correlated with conventional
 sources of cosmic
 rays
 and are consistent with an isotropic distribution \cite{sb}.
 This is consistent with the 
idea that
 they originate from decaying massive particles \cite{dp}.
 The particles must have a mass $
m_{X} \approx 10^{12} \GeV$ and lifetime $\tau_{X}
 \approx 10^{16} (\xi_{X}/3 \times
 10^{-4})\;yr$, where  $\xi_{X} \approx
 \Omega_{X}/\Omega_{\chi}$ is the fraction of
 halo
 CDM in the form of $X$ particles. For masses $10^{11} \GeV 
\lae m_{X} \lae 10^{13}
 \GeV$ there is no conflict with the diffuse 
$\gamma$-ray background or positron flux in
cosmic rays \cite{sb}. Therefore, as far as $X$ particles
 generated during inflation are
 concerned, the only restriction is that $\Omega_{X} < 0.3$. 

\section{Non-Thermal WIMP Dark Matter}

       The conditions in Section 2 are quite general and apply to dark matter
 particles regardless of their interaction with
 particles in the thermal background. For the more specific case of 
$Y$ particles decaying to WIMP dark matter and to
a significant number of charged particles or photons, we must impose two 
other conditions. 

             The first condition is that the WIMPs from $Y$ decay 
do not lose too
 much energy by scattering off of thermal background 
particles, since it is assumed
 that the WIMPs evolve by freely expanding after $Y$ decay.

        The condition for the WIMPs not to scatter is that, for the case of  
relativistic WIMPs, the scattering rate for WIMPs from thermal background 
particles should satisfy $\Gamma_{sc}
 \equiv n(T) \sigma_{sc} \Delta E< E H$,
where $n(T) = \frac{1.2 g(T) T^{3}}{\pi^{2}}$
 is the number density of particles in the
 thermal background \cite{eu}, E is the energy of the
 WIMP and $\Delta E$ 
is the energy lost by the WIMP per scattering
 with a particle in the thermal
background. In the following we will consider
 WIMPs with masses of the order of
 $m_{W}$ and unsuppressed weak
 interactions, such that the scattering with thermal background electrons and
 neutrinos is via t-channel $Z^{o}$ exchange.
 Experimentally, Dirac fermions and scalars
are excluded as dark matter particle in this
 case \cite{dmex}, so we consider the case of
 Majorana fermion WIMPs. In the Appendix
 we give the scattering cross-section for the
 scattering of relativistic Majorana fermions
 from thermal background particles and discuss
 the energy transfer per scattering.

         The scattering cross-section times energy lost per scattering
integrated over centre-of-mass scattering angle
 for Majorana WIMPs is (Appendix) 
\be{e11}    \sigma_{sc} \Delta E  \approx 
\frac{2 \pi \alpha^{2} k_{s}}{E_{\tau}}, \;\;\; 
k_{s} = \left(log\left(\frac{4 E_{\chi} E_{\tau}}{m_{Z}^{2}}\right)
 -\frac{1}{4}\right)  \;\;\;;\;\; 4 E_{\chi}E_{\tau} > m_{\chi,\;Z}^{2}  ~,\ee 
and
\be{e11a}    \sigma_{sc} \Delta E  \approx 
\frac{32 \pi \alpha^{2} k^{'}_{s}
 E_{\chi}^{4}E_{\tau}^{3}}{m_{\chi}^{4} m_{Z}^{4}}, 
\;\;\;  k_{s}^{'} = \frac{40}{3}
\;\;\;;\;\; 4 E_{\chi}E_{\tau} < m_{\chi,\;Z}^{2}   ~.\ee 
Here $\alpha = g_{\chi}g_{\tau}/4 \pi
 \approx 0.006$ (with $g_{\chi}$ and $g_{\tau}$
 the couplings of $\chi$ and the thermal background
 particles to $Z^{o}$ (Appendix)) and $E_{\tau} \approx 3 T $ is 
the energy of the thermal background particles. 

     For the case where $4E_{\chi}E_{\tau} 
> m_{\chi, \; Z}^{2}$ ($T_{d} > 
 {\rm Max}(m_{Z}^{2}, \;m_{\chi}^{2})/(3 \beta m_{Y})$),
 no-scattering imposes the constraint
\be{e12} m_{Y} \gae 
\frac{4 \alpha^{2} g(T_{d}) k_{s} M_{Pl}}{3 \pi \beta k_{T_{d}}}  
\approx 9 \times 10^{15} \left(\frac{1}{\beta}\right)
\left(\frac{\alpha}{0.006}\right)^{2}
\left(\frac{g(T_{d})}{100}\right)^{1/2}
\left(\frac{k_{s}}{10}\right)
 \GeV  ~,\ee
where $ k_{T} = \left(\frac{4 \pi^{3} g(T)}{45}\right)^{1/2} $. 
This is ruled out if $Y$ particles are to be
 generated by quantum fluctuations during
 inflation, since particles with $m_{Y} >  H_{I} 
\approx 10^{13} \GeV$ cannot be generated during inflation \cite{grav},
where $H_{I}$ is the expansion rate
 during inflation \footnote{It may be possible to generate
 such heavy particles via preheating after inflation \cite{preh}}. 

        For $4E_{\chi}E_{\tau} < m_{\chi, \; Z}^{2}$
 ($T_{d} <  {\rm Min}(m_{Z}^{2}, \;m_{\chi}^{2})/(3 \beta m_{Y})$),
we obtain an upper bound on the decay temperature
\be{e12a}   T_{d} \lae 
\left(\frac{1}{864 k^{'}_{s} g(T)}\right)^{1/4}
\frac{k_{T_{d}}^{1/4} m_{Z}
 m_{\chi}}{\alpha^{1/2}
\beta^{3/4}m_{Y}^{3/4}M_{Pl}^{1/4}}
~.\ee
Numerically this gives
\be{e13}   T_{d} \lae \frac{5.2 \times 10^{-3}}{\beta^{3/4}}
\left(\frac{0.006}{\alpha}\right)^{1/2}
\left(\frac{m_{\chi}}{100 \GeV}\right)
\left(\frac{m_{Z}}{90 \GeV}\right)
\left(\frac{100 \GeV}{m_{Y}}\right)^{3/4}
\left(\frac{10}{g(T_{d})}\right)^{1/8}
 \GeV
~. \ee
(We keep $m_{Z}$ explicit in this in order to check the effect of
 increasing the mass scale of the
 gauge interaction.) 

             Combined with the free-streaming constraint \eq{e6}, we find 
that in general no-scattering requires that 
\be{e14} m_{Y} \lae 10.4
\left(\frac{0.006}{\alpha}\right)^{2/7}
\left(\frac{1}{\beta}\right)
\left(\frac{r_{c}}{10^{-7}}\right)^{4/7}
\left(\frac{m_{\chi}}{100 \GeV}\right)^{8/7} 
\left(\frac{m_{Z}}{90 \GeV}\right)^{4/7}
\left(\frac{g(T_{d})}{10}\right)^{5/42}
\TeV  ~\ee
and
\be{e14a} T_{d} \lae 1.6 \times 10^{-4} 
\left(\frac{0.006}{\alpha}\right)^{2/7}
\left(\frac{10^{-7}}{r_{c}}\right)^{3/7}
\left(\frac{m_{\chi}}{100 \GeV}\right)^{1/7} 
\left(\frac{m_{Z}}{90 \GeV}\right)^{4/7}
\left(\frac{10}{g(T_{d})}\right)^{3/14}
\GeV   ~.\ee
Thus in order to both evade scattering and act as a
 source for non-thermal WIMP dark matter, 
$m_{Y} \lae 10$TeV and $T_{d}
 \lae 2 \times 10^{-4}$GeV is necessary. 

            In the above we have imposed the
 condition that the WIMPs effectively
 lose no energy through scattering
 with thermal background particles. However, if
$T_{d} <  {\rm Min}(m_{Z}^{2}, \;m_{\chi}^{2})/(3 \beta m_{Y})$ 
then the rate of loss of energy decreases as the particles lose energy, 
since $\Gamma_{sc} \Delta E/E
 \propto E_{\chi}^{3}$. So in this case, if $T_{d}$ 
is larger than the upper bound in \eq{e14a}, it
 may appear that it is possible for the 
WIMPs to lose energy and stop
 scattering from the thermal background
 whilst still remaining non-thermal, so allowing
 $T_{d}$ to evade the upper bound \eq{e14a} from no-scattering.
 However, it is easy to see that
 this is not the case: the effect of WIMPs losing
 energy to the thermal background is to
 effectively reduce the initial energy
 $\beta m_{Y}$ at $T_{d}$ until the no-scattering
 condition is satisfied. However, the upper 
bound from no-scattering and free-streaming,
 \eq{e14a}, is $\beta$ independent, so 
reducing $\beta$ will not allow $T_{d}$ to evade the upper bound. 

                A second condition, which applies if $T_{d} \lae
 T_{nuc} \approx 1$MeV and if there is a significant number of photons
 produced in the $Y$ decay cascade, is
 that the decaying $Y$ particles should not dissociate
 light elements produced during nucleosynthesis \cite{sarkar}. We
 will impose the constraint following from the non-dissociation of deuterium
 $D$, using the conservative bound based on
 assuming that a fraction $f_{\gamma}$ of the energy
 from $Y$ decays is entirely in the
 form of threshold 2.3MeV
 photons, which is the maximum number possible
 from the cascade decay \cite{lind}. 
$D$ non-dissociation then requires that
\be{e8} n_{2.3} \; p(2.3 \MeV) < n_{D}   ~,\ee
where $n_{2.3} \equiv f_{\gamma} 
\rho_{Y}(T_{d})/2.3 MeV$ is the number density of $2.3$MeV photons, 
$n_{D}$ is the number
 density of $D$ nuclei and $p(E)$ is the probability of a 
photon of energy $E$ photo-dissociating a $D$ nucleus \cite{lind}
\be{e9} p(E) = \frac{n_{D} \sigma_{D}}{n_{e} \sigma_{T}}     ~.\ee
$\sigma_{D}$ is the $D$ photo-dissociation cross-section, 
$\sigma_{T}$ is the total
 photon scattering cross-section from background particles 
(for threshold photons these
 are given by $\sigma_{D} = 3$mb and $\sigma_{T} = 125$mb \cite{lind}) 
and $n_{e}$ is the electron
 density. Using $n_{e} = n_{B} = \eta_{B} s(T)$,
 where $n_{B}$ is the baryon number density, $\eta_{B}$
 is the baryon number to entropy ratio and
 $s(T) \equiv \frac{2 \pi^{2}g(T)T^{3}}{45}$ is the entropy density,   
this results in a $T_{d}$-independent bound on the $Y$ mass
\be{e10} m_{Y} < 2.3 \; \frac{2 \pi^{2} g(T_{\gamma}) 
T_{\gamma}^{3}}{45} 
 \frac{\sigma_{T}}{\sigma_{D}}  \frac{\epsilon m_{\chi}
\eta_{B}}{f_{\gamma} \Omega_{\chi} \rho_{c}} \MeV   ~.\ee
Numerically this gives
\be{e10} m_{Y} \lae 0.26 \; \frac{\epsilon}{f_{\gamma} h^{2}} 
\left(\frac{\eta_{B}}{5 \times 10^{-11}}\right)
\left(\frac{0.3}{\Omega_{\chi}}\right)    
\left(\frac{m_{\chi}}{100 \GeV}\right) \GeV 
~,\ee
where critical density is 
$\rho_{c} = 7.5 \times 10^{-47} h^{2} \GeV^{4}$. 
However, since $\epsilon m_{\chi} \leq  m_{Y}$, this
 results in the condition 
\be{e10a} f_{\gamma} \lae \frac{2.6 \times 10^{-3}}{h^{2}}  
\left(\frac{\eta_{B}}{5 \times 10^{-11}}\right)
\left(\frac{0.3}{\Omega_{\chi}}\right)    ~.\ee
Therefore unless only a very small fraction of the $Y$ mass
 ends up in 
electromagnetic final states, deuterium dissociation rules out 
$Y$ decay after nucleosynthesis. 
  
      Taken in conjunction with the free-streaming and
 no-scattering constraint, which
 requires that $T_{d} \lae 2 \times 10^{-4} \GeV$, we
 see that in most cases the deuterium constraint
 rules out WIMPs with unsuppressed weak interactions 
as non-thermal dark matter. 
If we wish to retain WIMPs with unsuppressed
 weak interactions as dark matter 
we must have $Y$ particles decaying dominantly
 to WIMPs with little or no electromagnetic states produced in 
the decay. This is difficult, since in general
 a neutral weakly interacting particle will 
come in an $SU(2)_{L}$ representation together
 with electromagnetically charged particles. So we
 would require a model in which the mass of the Majorana WIMP
$\chi^{o}$ was lighter than its charged
 $SU(2)_{L}$ partners $\chi^{\pm}$, such that 
$2m_{\chi^{o}} < m_{Y} < 2m_{\chi^{\pm}}$. In
 this case $Y \rightarrow 2 \chi^{o}$
would be kinematically allowed but $Y \rightarrow \chi^{+}\chi^{-}$ disallowed.
Since the mass splitting of the $SU(2)_{L}$ representation must
 come from $SU(2)_{L}$ breaking and so must be
 at most of the order of $m_{W}$, the
 $Y$ mass must also be of the order of 
$m_{W}$ in any NTDM WIMP model with unsuppressed weak interactions. 

            Alternatively we can consider dark matter particles which interact 
more weakly with the thermal background than WIMPs with unsuppressed
weak interactions 
 and raise the $T_{d}$ upper bound above $T_{nuc}$. This can be done 
 either by reducing the strength of the WIMP
 coupling to the exchanged particle 
or by increasing the mass scale of the exchanged
 particle. From \eq{e14a} we see that 
this requires that $(m_{Z}/\alpha^{1/2})$ is increased by a factor of about 25
 to have an upper bound on $T_{d}$ greater than 1MeV, 
where we
 consider $m_{Z}$ to represent the mass scale of a general exchanged particle. 
From the free-streaming condition \eq{e6} we see that if $T_{d} \gae
 1$MeV then the decaying Y particle must have a mass greater than 
$40$TeV for $m_{\chi}$ of the order of
 $100 \GeV$, or more generally that 
$m_{Y}/m_{\chi} \gae 400$.  

 In \cite{hisano} 
it has been suggested that dark matter 
neutralinos could act as non-thermal dark matter. 
They find that for the case of neutralinos with unsuppressed $Z^{o}$ couplings
 (Wino limit), NTDM is ruled out, in agreement 
with the results derived here, but that for 
neutralinos with suppressed $Z^{o}$ couplings 
(Bino limit), which scatter from the
 thermal background via slepton exchange, 
and with heavy sleptons of mass of the order
 of 1TeV, it is marginally possible to have non-thermal
 neutralino dark matter with a free-streaming length of 0.1Mpc.

\section{Particle Densities From Quantum Fluctuations}

            In the previous sections we have 
considered the conditions for the dark matter
 particles from $Y$ decay to account for 
non-thermal dark matter. We now consider the
 origin of the heavy particles which may be responsible for NTDM and UHECR. 
In the following we will consider $\chi$ and $Y$ 
particles without imposing restrictions coming from their interactions
 with conventional matter. 
For example, it is possible that the non-thermal dark matter particles and
 decaying $Y$ particles could
 belong to a hidden sector which couples 
only very weakly if at all to conventional matter.
In this case $T_{d}$ could take any value without disrupting nucleosynthesis. 

               We will make the simplest assumptions regarding
 inflation, namely that the 
expansion rate $H$ is fixed during inflation, with inflation
 followed by a coherently oscillating inflaton matter dominated
 period characterized by a reheating temperature $T_{R}$. 

\subsection{$X$ and $Y$ Scalars} 

      We will consider the amplitude of the quantum fluctuations
 to be such that 
only the mass term in the scalar potential plays a role i.e. we
 neglect self-interactions. 
We also consider the case where $\phi = 0$ at the beginning of
 inflation (where $\phi$ represents the $X$ or $Y$ scalars),
 so that there is no
 significant contribution to the number density of scalars
 from an initial classical
 expectation value for $\phi$. 
For $m < H_{I}$, the scalars may be considered
 massless during inflation. 

                During each interval $\delta t \approx H_{I}^{-1}$, 
the scalar field receives a quantum
  fluctuation on horizon scales $\delta \phi \approx 
H_{I}/2 \pi$, which is thereafter 
 stretched beyond the horizon as a classical 
fluctuation. Therefore after $N$ 
e-foldings of inflation, the mean squared magnitude of the
 classical scalar
 field as seen on sub-horizon scales will be \cite{grav}
\be{e25} \delta \phi^{2}  
\approx \frac{N H_{I}^{2}}{4 \pi^{2}}       ~\ee
as a result of the random walk of the scalar field due to
 quantum fluctuations.
Once inflation ends and the Universe enters the inflaton
 matter dominated regime, 
the super-horizon wavelength modes begin to re-enter the horizon.
 We denote the expansion rate when the fluctuation of
 wavelength $\lambda$ re-enters the
 horizon as
 $H_{\lambda}$. The evolution of the 
modes then depends on whether they re-enter before or after $H = m$.
The modes obey the equation of motion 
\be{eq1} \delta \ddot{\phi} + 3 H \delta \dot{\phi}
 - \vec{k}^{2} \delta  \phi = -m^{2} \delta \phi    ~,\ee
where $\vec{k}$ is the wavenumber of the mode.
 Modes entering at
 $H_{\lambda} \leq m$ have $\delta \phi$ 
constant until $H = m$, after which they
 oscillate about $\delta \phi = 0$ with $\delta \phi \propto 1/a^{3/2}$. 
Modes entering at $H_{\lambda}
 > m$ are additionally suppressed, since their equation of motion is initially 
dominated by the
 $\vec{k}^{2}$ term. As a result their energy density evolves as 
$\delta \phi \propto 1/a$ until $\vec{k}^{2} 
\propto a^{-2}  < m^{2}$, after which the 
mass term dominates and the modes evolve as $\delta \phi \propto 1/a^{3/2}$. 
The average number
 density
 of $\phi$ particles at $H < m$ will therefore be given by 
\be{e26} n \approx m <\delta \phi^{2}> = 
n_{\left(H_{\lambda} < m\right)} 
+ n_{\left(H_{\lambda} > m\right)}= (N_{T}-N_{m})
 m \delta \phi^{2}_{c}
 + m <\delta \phi_{\left(H_{\lambda} > m \right)}^{2}>    ~,\ee
where $<...>$ denotes spatial average, 
$\delta \phi^{2}_{\left(H_{\lambda} > m\right)}$
 is the total contribution of modes with $H_{\lambda} > m$, 
$N_{m}$ is the 
number of e-foldings before the
 end of inflation at which modes 
re-entering at $H = m$ leave the horizon, $N_{T}$ is the total number 
of e-foldings of
 inflation ($N_{T} \gae 65$) and $\delta \phi_{c}$ is the 
amplitude of each $H_{\lambda}
 < m$ mode. In general, a mode re-entering at $H$ during 
inflaton matter domination 
exits the horizon during inflation at 
\be{e27}     N(H) = \frac{1}{3} ln\left( \frac{H_{I}}{H} \right)      ~.\ee
With $H_{I} \approx 10^{13} \GeV$ and $H = m \gae 
100 \GeV$, for example, 
$N_{m} \lae 8$. Thus to a reasonable approximation we 
can ignore the contribution of modes with $H_{\lambda} > m$
 and simply consider $N_{m} = 0$ in \eq{e26}. In this case the
 gradient terms in the $\delta \phi$ equation of motion 
can be neglected and all modes 
simply oscillate about the minimum of the potential such that 
 $\delta \phi \propto a^{-3/2}$ once $H \leq m$, 
whlist remaining constant at $H > m$.  
Therefore during matter domination by the inflaton ($H \propto a^{-3/2}$), 
$\delta \phi_{c}$ is given by
\be{e28}   \delta \phi_{e} \approx \frac{H}{m} 
\frac{H_{I}}{2 \pi}   ~.\ee
Thus
during inflaton matter domination, the number density of 
$\phi$ scalars is
\be{e29}   n(H) \approx \frac{N_{T}}{4 \pi^{2}} 
\frac{H^{2} H_{I}^{2}}{m}   ~.\ee
The number density at temperatures $T \leq T_{R}$ is then 
\be{e30}  n(T) \approx \frac{N_{T} \pi g(T)}{45} 
\frac{T_{R} T^{3} H_{I}^{2}}{m M_{Pl}^{2}}    ~.\ee

\subsection{$X$ and $Y$ Fermions}

           Quantum production of fermions during inflation 
is possible only if the fermions are massive \cite{gferm}. 
The resulting density during matter domination
 has been
 estimated to be \cite{gferm} 
\be{e31} n = C_{\alpha} m H^{2}     ~,\ee
where $C_{\alpha} \approx 3 \times 10^{-3}$ 
for a transition from inflation to matter
 domination. In general, the density of scalars will be 
much larger than that of fermions of the same mass, by a factor 
\be{e31a} \frac{N_{T}}{4 \pi^{2} 
C_{\alpha}}\left(\frac{H_{I}}{m}\right)^{2} \gg 1   ~.\ee 
As a result, if one has scalars and fermions with similar
 decay rates to dark matter particles (for example, supersymmetric
 partners), then the number of dark matter particles produced will be
 determined by the scalars.

\subsection{NTDM from Scalars}

       In order to account for NTDM from the decay of $Y$ scalars 
we require that 
$\Omega_{\chi} \equiv \epsilon n(T_{\gamma})m_{\chi}/\rho_{c}= 0.3$. Using
 \eq{e30} we find that this requires that
\be{e32}   T_{R} \approx \frac{45 m_{Y} M_{Pl}^{2} 
\rho_{c} \Omega_{\chi}}{\epsilon m_{\chi} 
N_{T} \pi H_{I}^{2}
 g(T_{\gamma}) T_{\gamma}^{3}}    ~.\ee 
Numerically this gives
\be{e32a} T_{R} \approx 280 \;
\left(\frac{60}{N_{T}}\right)
\left(\frac{h^{2}}{\epsilon}\right)
\left(\frac{m_{Y}}{100 \GeV}\right)
\left(\frac{100 \GeV}{m_{\chi}}\right)
\left(\frac{10^{13} \GeV}{H_{I}}\right)^{2}
\left(\frac{\Omega_{\chi}}{0.3}\right) \GeV  ~.\ee

\subsection{UHECR from Scalars}

                The requirement that $\Omega_{X} < 0.3$
 implies the $m_{X}$-independent
constraint
\be{e35}   T_{R} \lae 0.3 \frac{45 M_{Pl}^{2} 
\rho_{c}}{N_{T} \pi H_{I}^{2}
 g(T_{\gamma}) T_{\gamma}^{3}}    ~.\ee 
Numerically this gives
\be{e36}   T_{R} \lae 280 h^{2} 
\left(\frac{10^{13} \GeV}{H_{I}}\right)^{2}
\left(\frac{60}{N_{T}}\right) \GeV    ~.\ee
Thus a low reheating temperature is necessary 
in order to allow UHECR to be
 explained by scalars generated by quantum fluctuations.

\subsection{Relationship Between $X$ and $\chi$ Densities from Scalars}

              It is important to note that the mass density of $X$ 
and $Y$ scalars from
 quantum fluctuations is approximately the same, 
since the number density \eq{e30} 
is inversely proportional to $m$. (This is in contrast with
 the case of fermions, where the number density is proportional to $m$.)
Therefore, in general, the present 
density of dark matter particles from $Y$ scalar decay is related to the 
density of $X$ scalars by 
\be{e36a}  \frac{\Omega_{\chi}}{\Omega_{X}} = f_{\Omega}
 \frac{\epsilon m_{\chi}}{m_{Y}}      ~, \ee
where $f_{\Omega}$, as discussed below, parameterizes the uncertainty in
the number density of scalars from quantum fluctuations.
Since $\epsilon m_{\chi} \leq m_{Y}$, in order to have $\Omega_{\chi} 
> \Omega_{X}$ and so have dark matter primarily in the form of
 non-thermal $\chi$
 dark matter, we must have
 $f_{\Omega} > 1$. If $\Omega_{\chi} > \Omega_{X}$ then 
$m_{Y}/\epsilon \geq m_{\chi} > 
 m_{Y}/f_{\Omega}\epsilon$, and so for $\epsilon$ and $f_{\Omega}$ 
not very different from 1 $m_{\chi}$ and $m_{Y}$
 must be of the same order of
 magnitude. 

$f_{\Omega}$  parameterizes the fact that the number
 density \eq{e30} has an uncertainty. For example, 
if the total number of e-foldings of
 inflation $N_{T}$ is much larger than 65, then 
most of the number density in $Y$ scalars
 would be due to superhorizon modes, 
resulting in a coherently oscillating field $\delta \phi(t)$ in
 the Universe. The initial amplitude of this field is determined
 by a 1-dimensional
 random walk of the field during inflation due to 
quantum fluctuations. For $n$ steps of size 
$\pm a$, the probability of a particle being at $x$
 is described by the normal distribution
\be{e36c} P(x)dx =  
\frac{e^{-\frac{x^{2}}{2 a^{2} n}}dx}{(2 a^{2} n \pi)^{1/2}}
~.\ee
The root mean square value is given by $x_{rms}^{2} = n a^{2}$.
In our case $x \equiv \delta \phi$, $a \equiv H_{I}/(2 \pi)$ and 
$n \equiv N_{T}$. As an estimate of the range of values $f_{\Omega}$ can 
reasonably take
we integrate \eq{e36c} and exclude the range of values $(0,|\delta
 \phi_{1}|)$ and 
$(|\delta \phi_{2}|, \infty)$ for which the probability is less
 than 0.316, chosen to give the range of values of $f_{\Omega}$
 which have a probability of 0.8. This gives 
$|\delta \phi_{1}| = 0.41 |\delta \phi_{rms}|$ and $|\delta \phi_{2}| 
= |\delta \phi_{rms}|$. 
(Note that the best estimate of the mean value of $|\delta \phi|$ is 
somewhat smaller 
than $|\delta \phi_{rms}|$. If we take the value
 for which the probability of being larger
 or smaller is 0.5, then the mean value is $0.7 |\delta \phi_{rms}|$.) 
The number density $n \propto \delta \phi^{2}$ can
 therefore reasonably take values in the range 0.34 to 2.04 
 times the mean density. 
So in this case $f_{\Omega} = n_{Y}/n_{X}$
 can reasonably be in the range 0.17 to 5.9, with the
 probability of being smaller or larger than
 these values being $0.316^{2} = 0.1$. (This is
 independent of the definition of the mean density.)
 So long as $f_{\Omega} > 1$, the dark matter density can be primarily
 due to non-thermal $\chi$ dark matter. 
 However, we still expect to find a significant
 contribution to the total dark matter density
 from $X$ scalars. This is important, since in
 this case galactic halos will be composed
 of a combination of free streaming non-thermal dark matter
 particles and a smaller but possibly dynamically significant 
component of conventional
 cold dark matter due to $X$ scalars.
 This may result in different predictions for
 halo formation and small-scale structure as
 compared with the limiting cases of pure non-thermal dark matter
 or pure cold dark matter. 

\subsection{NTDM from Fermions}

     In this case, in order to account for $\Omega_{\chi} = 0.3$ we require that
\be{e37} T_{R}  =  
\frac{45 M_{Pl}^{2} \rho_{c} \Omega_{\chi}}{4 \pi^{3} g(T_{\gamma}) 
C_{\alpha} T_{\gamma}^{3} \epsilon  m_{\chi} m_{Y}}    ~.\ee 
Numerically this gives,
\be{e34a} T_{R} = 4.3 \times 10^{24} \;
\frac{h^{2}}{\epsilon C_{\alpha}}
\left(\frac{100 \GeV}{m_{\chi}}\right)
\left(\frac{100 \GeV}{m_{Y}}\right)
\left(\frac{\Omega_{\chi}}{0.3}\right)
\GeV  ~.\ee
Since $m_{Y} \lae H_{I} \approx 10^{13} \GeV$ for fermions
 generated from quantum fluctuations, there is a lower bound on the reheating
temperature as a function of $m_{Y}$, 
\be{e34b} T_{R} \gae  1.4 \times 10^{16}\;  \frac{h^{2}}{\epsilon} 
\left(\frac{3 \times 10^{-3}}{C_{\alpha}}\right)
\left(\frac{100 \GeV}{m_{\chi}}\right)
\left(\frac{\Omega_{\chi}}{0.3}\right)
\GeV    ~.\ee
For example, we consider two mass scales for $\chi$ and $Y$ particles
 which are of particular interest: 
the weak scale $m_{W}$ and the mass scale $10^{12-13} \GeV$
 associated with $X$ particles and $H_{I}$.
Since the largest possible reheating temperature 
after inflation is $T_{R} \approx 
(H_{I} M_{Pl})^{1/2} \approx 10^{16} \GeV$, the case $m_{\chi} 
\sim m_{W}$ is only marginally compatible with $Y$ fermions from quantum
 fluctuations and only if $m_{Y} \approx 10^{13} \GeV$.  The 
free-streaming constraint,
\eq{e6}, then implies that the $Y$ fermions decay at temperature
\be{e34c} T_{d} \approx 2.5 \times 10^{5} 
\left(\frac{10^{-7} \beta}{r_{c}}\right)
\left(\frac{g(T_{\gamma})}{g(T_{d})}\right)^{1/3}
\left(\frac{100 \GeV}{m_{\chi}}\right)
\left(\frac{m_{Y}}{10^{13} \GeV}\right) \GeV
~.\ee
(In this case WIMPs are ruled out as dark matter 
by the no-scattering constraint, \eq{e12}.)
For the case of very large $\chi$ mass, $m_{\chi} \approx 10^{13} \GeV$,
 the reheating temperature must satisfy
\be{e34d} T_{R} \gae  1.4 \times 10^{5} \; \frac{h^{2}}{\epsilon} 
\left(\frac{3 \times 10^{-3}}{C_{\alpha}}\right)
\left(\frac{10^{13} \GeV}{m_{\chi}}\right)
\left(\frac{\Omega_{\chi}}{0.3}\right)
\GeV    ~,\ee
where we have imposed $m_{Y} \approx 10^{13} \GeV$ 
since $m_{Y} \geq \epsilon m_{\chi}$, 
whilst the $Y$ fermions decay at 
\be{e34e} T_{d} \approx 2.5 \times 10^{-6} 
\left(\frac{10^{-7} \beta}{r_{c}}\right)
\left(\frac{g(T_{\gamma})}{g(T_{d})}\right)^{1/3}
\left(\frac{m_{Y}}{m_{\chi}}\right) \GeV
~.\ee
(In this case WIMPs are ruled out as dark matter by the 
nucleosynthesis constraint.)

\subsection{UHECR from Fermions} 

              In this case the constraint $\Omega_{X} < 0.3$ implies that
\be{e37} T_{R}  < 0.3 \;
\frac{45 M_{Pl}^{2} \rho_{c}}{8 \pi^{3} C_{\alpha}
 T_{\gamma}^{3} m_{X}^{2}}    ~.\ee 
Numerically this gives
\be{e38}   T_{R} \lae 1.4 \times 10^{7} h^{2} 
\left(\frac{3 \times 10^{-3}}{C_{\alpha}}\right)
\left(\frac{10^{12} \GeV}{m_{X}}\right)^{2}
\GeV    ~.\ee
Thus a wide range of reheating temperatures is
 consistent with UHECR from $X$
 fermions. 

\section{Conclusions}

     We have considered the conditions under which non-singular 
galactic halos and UHECR can be explained by decaying particles produced 
during inflation by quantum fluctuations.
 For the case of WIMP non-thermal dark matter from 
decaying particles, the requirement that
 WIMPs with unsuppressed weak interactions 
do not lose energy by scattering from
 the thermal background combined with the
 requirement that their free-streaming length 
is of the order of 0.1Mpc implies that the $Y$ particles decay at temperatures 
less than that of nucleosynthesis. Thus unless $Y$ particles can decay to WIMPs
without producing a significant number of photons in the cascade, WIMPs 
with unsuppressed weak interactions are ruled out by photodissociation 
of deuterium. 
It may be marginally possible to suppress the weak interactions of the WIMPs
 sufficiently that the $Y$ particles can decay above 1 MeV, for example 
in the case of supersymmetry with Bino dark matter and with
 heavy sleptons of mass $\sim 1$TeV \cite{hisano}. 

 An alternative possibility is to consider the
 case where the dark matter particles and $Y$ 
particles belong to a sector interacting only
 very weakly if at all with conventional matter, 
such that there is no danger to nucleosynthesis from $Y$ decay products. 
In the case where UHECR and
NTDM originate simultaneously from
$X$ and $Y$ scalars generated by quantum fluctuations, 
the present mass density of $X$ scalars will naturally be of the same
 order of magnitude as that of the
 non-thermal dark matter. In this case we expect
 the dark matter in the galactic halo to be a combination
 of free-streaming non-thermal dark matter and
 conventional $X$ scalar cold dark
 matter, which could alter the predictions for galactic
 halo and small-scale
 structure formation from the case of pure non-thermal dark matter. 
This would allow an observational test of
 the idea that decaying scalar particles
 generated during inflation
can explain both UHECR and non-singular galactic halos. 

\section*{Appendix. Majorana Fermion Scattering Cross-Section}

           The scattering cross-section in the CM frame is 
\be{ap1} \left(\frac{d \sigma}{d \Omega}\right)_{CM} = \frac{1}{64 \pi^{2} s}
 \frac{|\vec{p}_{f}|}{|\vec{p}_{i}|} |\overline{{\cal M}}|^{2}   ~,\ee
where $\vec{p}_{i,\;f}$ are the initial and final three-momenta and $s$ is the
CM energy squared. 
We consider a Majorana fermion $\chi$ of mass $m_{\chi}$ scattering from 
 a massless thermal background fermion $\tau$.
 For simplicity we will consider a head-on
 collision of $\chi$ with $\tau$ along
 the x-direction; this can generally be arranged 
by boosting the frame in the cases
 where the collision is at an angle, resulting in 
shifts in the energy and momenta by factors
 of the order of 1. The weak interactions are 
described by 
\be{ap2} {\cal L}_{int} = 
\frac{g_{\chi}}{2} \overline{\chi}
 \gamma^{\mu} \gamma_{5} \chi Z_{\mu} 
+ g_{\tau} \overline{\tau}
 \gamma^{\mu} (1 - \gamma_{5})\tau Z_{\mu}
~.\ee
In numerical estimates  we use for $g_{\chi}$ and $g_{\tau}$ 
the value of the $\nu_{L} \nu_{L} Z^{o}$ 
coupling from the Standard Model, $g_{\tau}
 = g{\chi}/2  = g_{2}/4 Cos \theta_{W} = 
0.19$, where $g_{2}$ is the $SU(2)_{L}$ gauge coupling.
 The spin-averaged amplitude squared
 for t-channel $Z^{o}$ exchange is then
\be{ap3}  |\overline{{\cal M}}|^{2}
 = \frac{16 g^{2} g^{'\;2}}{(k^{2} -
 m_{Z}^{2})^{2}} [(p_{1}.q_{1})(p_{2}.q_{2})
 + (p_{1}.q_{2})(p_{2}.q_{1}) 
+ m_{\chi}^{2} (q_{1}.q_{2}) ]     ~,\ee
where $p_{1,2}$ are the initial and final $\chi$ 4-momenta, 
$q_{1,2}$ are the initial and final $\tau$ 4-momenta and 
$k^{2} = (q_{2}- q_{1})^{2}$. 
For $E_{\chi} \gg E_{\tau},\; m_{\chi}$ we find 
\be{ap4}  |\overline{{\cal M}}|^{2} = \frac{16 g^{2} g^{'\;2}}{(k^{2} -
 m_{Z}^{2})^{2}} 
\frac{8 E_{\tau}^{4} \gamma^{4}}{\eta^{2}} 
[1 + 2 \eta (1+Cos\theta) + 2 \eta^{2}(1 + Cos^{2}\theta)
+ \frac{m_{\chi}^{2}}{2 E_{\tau}^{2}}
\frac{\eta^{2}}{\gamma^{2}}(1-Cos \theta) ] 
     ~,\ee
where $\theta$ is the scattering angle in the CM frame, 
$\eta$ and $\gamma$ are defined by
\be{ap5} \eta^{2} = \frac{E_{\chi}^{2} E_{\tau}^{2}}{(m_{\chi}^{2}
+2 E_{\chi}E_{\tau})^{2}},\;\;\;\;\;\
\gamma^{2} = \frac{E_{\chi}^{2}}{(m_{\chi}^{2}
+4 E_{\chi}E_{\tau})}   ~,\ee
and 
\be{ap6} k^{2} = 
\frac{-8 E_{\chi}^{2}E_{\tau}^{2}(1-Cos\theta)}{m_{\chi}^{2}+ 
4 E_{\chi}E_{\tau}}   ~.\ee
Also $s = m_{\chi}^{2} + 4 E_{\chi} E_{\tau}$. 
The energy loss per scattering in the LAB frame (the
rest frame of the thermal background)
for a Majorana fermion scattering
at angle $\theta$ in the CM frame from a thermal background
 particle, in the limit $E_{\chi} \gg m_{\chi}, E_{\tau}$, is
given by
\be{ap9} \Delta E =  
\frac{2 E_{\chi}^{2}
 E_{\tau}(1-Cos\;\theta)}{(m_{\chi}^{2} + 4 E_{\chi}E_{\tau})}
~.\ee 

                   In the limit $ 4 E_{\chi} E_{\tau} > m_{\chi,\;Z}^{2}$ the
cross-section times energy lost per scattering integrated over
 scattering angle is then
\be{ap7} \sigma_{sc} \Delta E 
\equiv \int d\Omega \left(\frac{d \sigma}{d \Omega}\right)_{CM} \Delta E(\theta)
 \; \approx \frac{ g^{2} g^{'\;2}}{8 \pi}
 \frac{k_{s}}{E_{\tau}}\;\;;\;\;\; 
k_{s} = \left(log\left(\frac{4 E_{\chi} E_{\tau}}{m_{Z}^{2}}\right) -
\frac{1}{4}\right)  
  ~.\ee 
In the limit $4 E_{\chi} E_{\tau}
 < m_{\chi,\;Z}^{2}$ the cross-section times energy lost per scattering is
\be{ap8} \sigma_{sc} \Delta E 
\approx \frac{ 2 g^{2} g^{'\;2}}{\pi}
 \frac{k^{'}_{s}E_{\chi}^{4}
E_{\tau}^{3}}{m_{\chi}^{4}m_{Z}^{4}}, \;\;\; ;\;\;\;
 k^{'}_{s} = 40/3    ~.\ee

\end{document}